\documentclass[letterpaper,12pt,preprint]{aastex}

\usepackage{amssymb,amsmath,amsbsy}

\usepackage{color,hyperref}
\definecolor{linkcolor}{rgb}{0,0,0.5}
\hypersetup{colorlinks=true,linkcolor=linkcolor,citecolor=linkcolor,
                    filecolor=linkcolor}
\usepackage{url}

\newcommand{\given}{\,|\,}
\newcommand{\dd}{\mathrm{d}}

\newcommand{\msun}{\mathrm{M}_\odot}

\begin{document}

\title{A re-interpretation of the Triangulum-Andromeda stellar clouds: a population of halo stars kicked out of the Galactic disk}
\author{Adrian M. Price-Whelan\altaffilmark{\colum}, Kathryn V. Johnston\altaffilmark{\colum},
	    Allyson A. Sheffield\altaffilmark{\lga}, Chervin F. P. Laporte\altaffilmark{\colum}, 
	    Branimir Sesar\altaffilmark{\mpia}}

\newcommand{\colum}{1}
\newcommand{\lga}{2}
\newcommand{\mpia}{3}
\altaffiltext{\colum}{Department of Astronomy, 
		              Columbia University, 
		              550 W 120th St., 
		              New York, NY 10027, USA}
\altaffiltext{\lga}{LaGuardia Community College,
			City University of New York,
			Department of Natural Sciences,
			31-10 Thomson Ave.,
			Long Island City, NY 11101, USA}
\altaffiltext{\mpia}{Max-Planck-Institut f\"ur Astronomie,
                     K\"onigstuhl 17, D-69117 Heidelberg, Germany}

\begin{abstract}
The Triangulum-Andromeda stellar clouds (TriAnd1 and TriAnd2) are a pair of concentric ring- or shell-like over-densities at large $R$ ($\approx$ 30 kpc) and $Z$ ($\approx$ -10 kpc) in the Galactic halo that are thought to have been formed from the accretion and disruption of a satellite galaxy.
This paper critically re-examines this formation scenario by comparing the number ratio of RR Lyrae to M giant stars associated with the TriAnd clouds with other structures in the Galaxy. 
The current data suggest a stellar population for these over-densities ($f_{\rm RR:MG} < 0.38$ at 95\% confidence) quite unlike any of the known satellites of the Milky Way ($f_{\rm RR:MG} \approx 0.5$ for the very largest and $f_{\rm RR:MG} >>1$ for the smaller satellites) and more like the population of stars born in the much deeper potential well inhabited by the Galactic disk ($f_{\rm RR:MG} < 0.01$).
N-body simulations of a Milky-Way-like galaxy perturbed by the impact of a dwarf galaxy demonstrate that, in the right circumstances,  concentric rings propagating outwards from that Galactic disk can plausibly produce similar over-densities.
These results provide dramatic support for the recent proposal by \citet{xu15} that, rather than stars {\it accreted} from other galaxies, the TriAnd clouds could represent stars {\it kicked-out} from our own disk.
If so, these would be the first populations of disk stars to be found in the Galactic halo and a clear signature of the importance of this second formation mechanism for stellar halos more generally.
Moreover, their existence at the very extremities of the disk places strong constraints on the nature of the interaction that formed them.
\end{abstract}


\section{Introduction}\label{sec:introduction}

Formation scenarios that have been proposed for the diffuse stellar halos that surround galaxies can be broadly divided into three categories.
First, stars might be  {\it accreted} from other dark matter halos that have been disrupted by the main galaxy \citep[e.g.,][]{searle78}.
The last decade has revealed abundant evidence for this mechanism in the form of debris structures in space and velocity around our own \citep[e.g.][]{majewski03,belokurov06,schlaufman09}  and other \cite[e.g. M31, see][]{ibata07,gilbert09} galaxies.
Second, stars might form {\it in situ} from gas located in the the halo itself \citep[e.g.][]{els62}.
While such populations have been seen in hydrodynamic simulations of galaxy formation \citep{tissera12} it is as yet unclear whether their formation is an artifact of either numerical techniques or the assumed star-formation prescriptions. The observational evidence for  the existence of this population remains controversial \citep[e.g.][for opposing views]{carollo08,schoenrich14}.
Lastly, stars initially made within a galaxy's disk might be {\it kicked-out} onto halo orbits by some perturbation \citep{zolotov09, zolotov10, mccarthy12}.
Deep star count surveys of M31 have revealed  a disturbed disk-like structure extending to 40--50 kpc from the center of the galaxy \citep{ferguson02} which contain stellar populations \citep{richardson08,bernard15} indicating that kicked-out stars do indeed contribute to its inner halo \citep[and see][for comparison simulations]{purcell10}. However, observations of {\it kicked-out} stars around our own Galaxy are at this point suggestive rather than conclusive \citep[e.g.][]{sheffield12,hawkins15}

\begin{figure}[p]
\begin{center}
\includegraphics[width=\columnwidth]{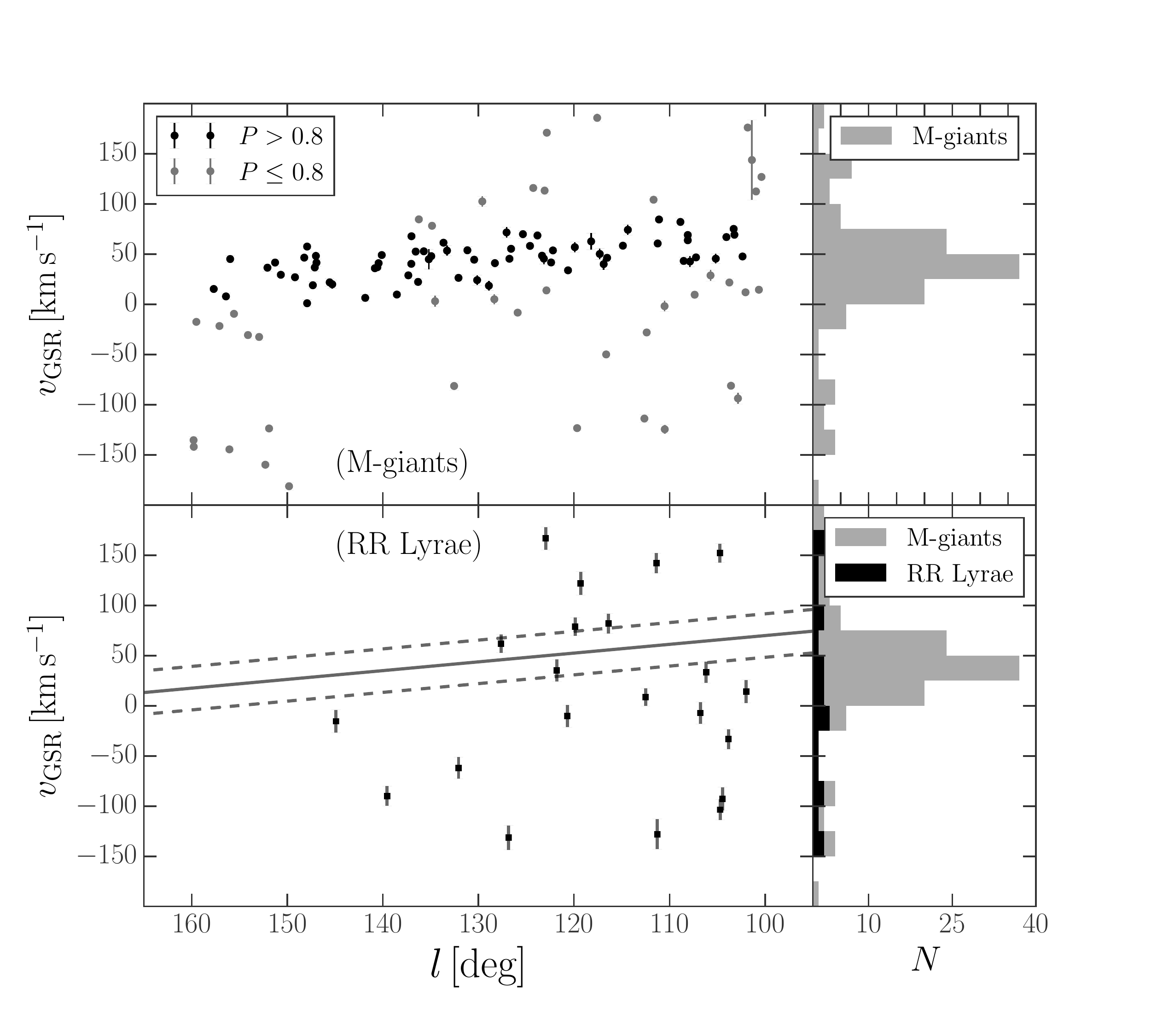}
\caption{\label{fig:data} Galactocentric radial velocities as a function of Galactic longitude, $l$, for M giants and RR Lyraes in the TriAnd region. Top left panel shows the velocities of all M giants in the TriAnd region: black circles are stars with $>$80\% probability of belonging to TriAnd, grey circles have $\leq$80\% probability of belonging to TriAnd and are probably typical halo stars. Velocity uncertainties are typically around the size of the marker. Top right panel shows a 1D histogram of the M giant velocities. Bottom left panel shows a summary of the MAP fit of our velocity model (Section~\ref{sec:velmodel}) for the TriAnd sequence (solid grey line), plus or minus the derived MAP velocity dispersion of TriAnd (dashed lines). Over-plotted are the velocities of our sample of RR Lyrae stars in the TriAnd region. Bottom right shows the same M giant histogram (grey), with a histogram of the RR Lyrae velocities (black) with the same bins.}
\end{center}
\end{figure}

The Triangulum-Andromeda clouds are examples of stellar over-densities in the halo of the Milky Way thought to have formed through the disruption of a satellite and hence contributing to the {\it accreted} component of the stellar halo.
A single over-density was originally identified in M giant stars selected from the Two Micron All Sky Survey (2MASS) covering the region $100^\circ < l < 160^\circ$ and $-35^\circ < b <-15^\circ$ at approximately 20 kpc from the Sun \citep{rochapinto04}. 
Spectroscopic follow-up of these M giants showed them to have a velocity distribution with a small dispersion ($\approx$25 km~s$^{-1}$) and a shallow gradient with mean velocities increasing with decreasing $l$ (in the Galactic-Standard-of-Rest -- GSR -- frame). 
A {\it double} sequence of main-sequence stars  in the foreground of the PANDA's survey was later found in the same region  \citep[called TriAnd1 and TriAnd2 by][]{martin07},
and \citet{martin14} subsequently presented evidence for the presence of several additional structures.
In our own, recent study \citep{sheffield14}, we found that the 2MASS data suggests that there are M giant counterparts to both of the primary TriAnd1 and TriAnd2  sequences. 
Requiring both M giant and main-sequence stars to belong to the  same populations and fitting isochrones to both simultaneously gave (distance, age, [Fe/H]) estimates of ($\approx$18 kpc,6-8 Gyrs, -0.75) and ($\approx$28 kpc, 10-12 Gyrs, -0.85) respectively (i.e. Galactocentric $(R,Z)\approx(24,-8)$ kpc and $(33,-12)$ kpc for the central pointing at $(l,b)=(125^\circ,-25^\circ)$, although the center of the over-density itself could be hidden at lower $|z|$ behind the Galactic disk). 
Using a spectroscopic survey we showed that M giants in both TriAnd1 and TriAnd2 followed the same velocity distribution found in the original study (Figure~\ref{fig:data}, top panels).
We proposed that the large Galactocentric radius and distance below the mid-plane combined with the small dispersion was suggestive of neither a disk nor random halo population.
We used simulations of satellite disruption to show that both TriAnd1 and TriAnd2 could plausibly be due to single accretion event of a satellite on a retrograde orbit --- the two distinct structures overlapping in orbital phase, but corresponding to material lost on two separate pericentric passages.

In this paper we re-examine our proposal for the origin for TriAnd1 and TriAnd2  using the results from our own spectroscopic survey of RR Lyrae stars in the region. 
The original aim of this survey was to use the spectra to identify RR Lyrae that followed the same velocity sequence as the M giants and hence could be associated with these structures.
The properties of the associated RR Lyrae would provide better constraints on the distances to TriAnd1 and TriAnd2 as well as shed light on their stellar populations.

In Section \ref{sec:data} we describe our target selection, data collection and reduction. 
In Section \ref{sec:resultsi} we present the results of the survey and discuss the implications for the stellar populations (as represented by the number ratio of RR Lyrae to M giants stars, $f_{\rm RR:MG}$) present in the TriAnd clouds.
In Section \ref{sec:resultsii} we discuss the implications of our findings the origin of TriAnd1 and TriAnd2 and 
in Section \ref{sec:resultsiii} we present a numerical simulation to illustrate this possible formation mechnism.
We summarize our conclusions and outline future work in Section \ref{sec:conclusions}.

\section{Data} \label{sec:data}

TriAnd1 was originally identified as an over-density in M giant stars selected from the Two Micron All Sky Survey. Based on the stellar populations of Milky Way satellites and of other halo over-densities \citep[e.g., the Hercules-Aquila cloud;][]{simion14}, we also expect to find an over-density of RR Lyrae stars (RRLs) in this region. RR Lyrae stars have several properties that make them useful as tracers of halo over-densities. First, they are bright stars ($M_r=0.6$ mag at ${\rm [Fe/H]}=-1.5$ dex) that can be detected at large distances (5--120 kpc for $14 < r < 21$). Second, distances to RRLs measured from optical data are precise to $\approx6\%$ \citep{sesar13} versus, e.g., 15--20\% for M giants \citep{sheffield14}. Finally, RRab stars have distinct, saw-tooth shaped light curves which make them easy to identify given multi-epoch observations (peak-to-peak amplitudes close to 1 mag in the $r$-band and periods of $\approx0.6$ days). Our original goal was to obtain spectra of RRLs selected from the TriAnd region in order to identify stars with velocities consistent with the M giant population, thus allowing for an accurate measurement of the distance to the TriAnd structures. 


\subsection{Obtaining velocities for RR Lyrae in the TriAnd region}
\subsubsection{Target selection: PTF}

We selected RRLs from the Palomar Transient Factory \citep[PTF;][]{law09} in the region around TriAnd1 and TriAnd2: Galactic longitude and latitude, $(l,b)$, in the ranges $100^{\circ} < l < 160^{\circ}$, $-35^{\circ} < b < -15^{\circ}$, and heliocentric distances in the range $15 < d < 35~{\rm kpc}$. We follow the procedure of \cite{drake13} and use the absolute-magnitude-metallicity relation of \citet{catelan08} to compute the distances, where we assume all stars have a metallicity of ${\rm [Fe/H]}=-1.5$. There are 70 RR Lyrae detected in PTF that fall in this volume.

\subsubsection{Data collection and reduction}

We used the 2.4 meter Hiltner telescope at MDM Observatory on two observing runs (August 2013 and October 2013) to obtain spectra for 20 of the RR Lyrae stars selected from PTF. We used the Modular Spectrograph (ModSpec) with the 600 l mm$^{-1}$ grating blazed at 4752 \AA, centered near the prominent H$\alpha$ absorption feature, and the Templeton CCD. This set-up provides spectral coverage from 5300 \AA~ to 7200 \AA; this range was selected to focus on the H$\alpha$ line, which we use to derive the systemic velocities of the RR Lyrae stars. We found that total exposure times of 1200 seconds are adequate to achieve a signal-to-noise of ${\rm S/N} \approx 15$ near the H$\alpha$ line for an RR Lyrae with a $V$-band magnitude of 17. HgNe lamps were taken before and after each observation, and 2 sets of 10 flat field frames were taken, one set at the beginning of the night and the other set at the end. Heliocentric radial velocities for the H$\alpha$ line were found using standard IRAF tasks. The motion of the observer was corrected for using the \texttt{rvcorr} task. To check for systematic shifts in the wavelength calibration, we used the [O I] night sky emission line at 6300.304 \AA; no systematic shifts were found, although we did shift any spectra that had a deviation greater than 0.1 \AA~ so that the observed value of the line matched the lab value. 

\begin{figure}[p]
\begin{center}
\includegraphics[width=\columnwidth]{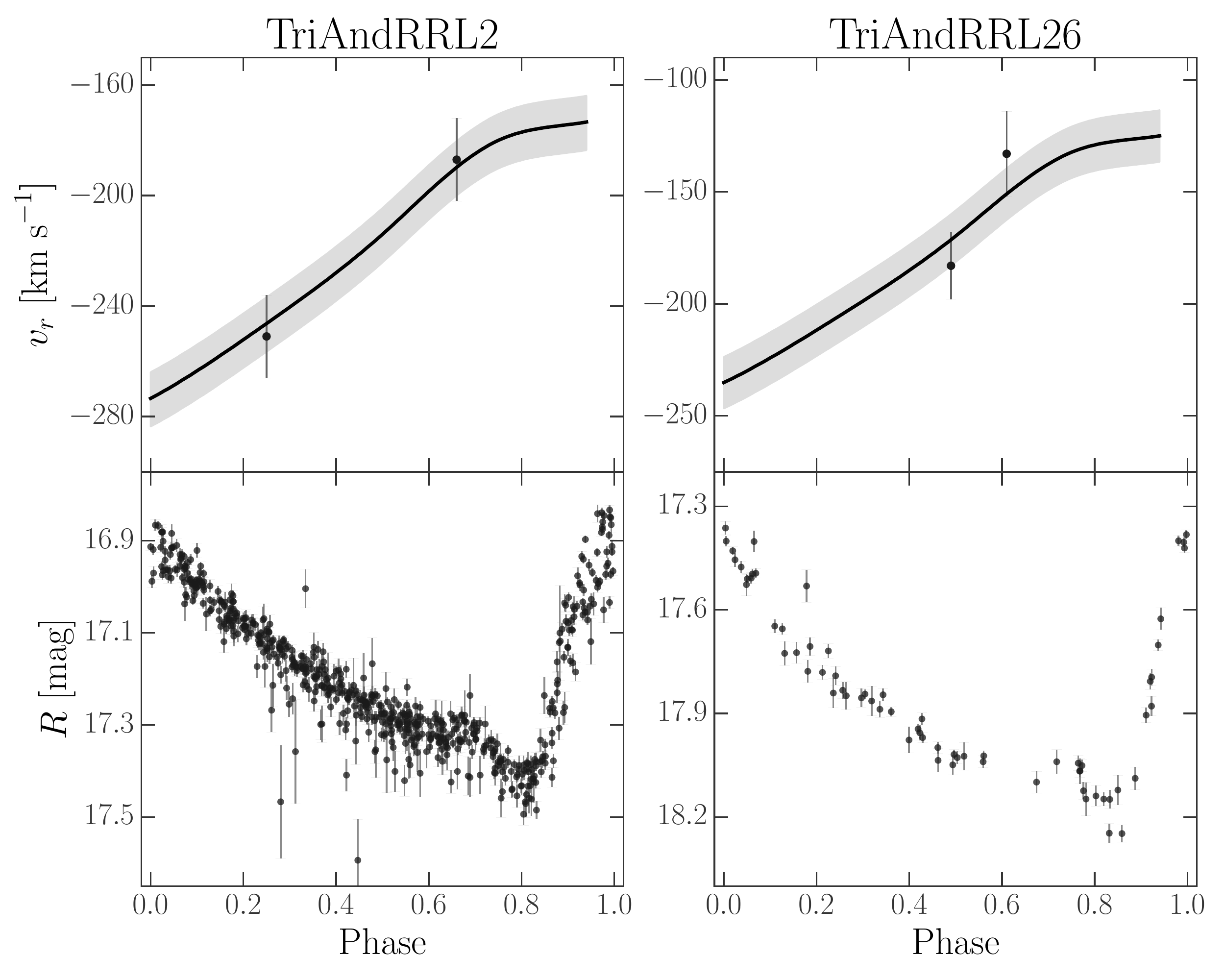}
\caption{Examples of template fits to radial velocity measurements of two of our sample of RR Lyrae stars (top). Black lines show the best-fit template. Grey, shaded region show the 68\% confidence region. Bottom panels show phased, $R$-band light curves for the two stars, demonstrating the data quality of PTF and two different extremes of number of observations.  \label{fig:rvcurve}}
\end{center}
\end{figure}

To derive the systemic velocities of the RR Lyrae stars, we use the procedure outlined in \citet{sesar12}. The systemic velocities are found using the H Balmer lines, although a different template is used for each line --- we use the template derived for the H$\alpha$ line. For most stars, we took spectra at 2 phase points within the phase range 0.05--0.85, however several stars have only 1 phase point measured. We use standard least-squares optimization to fit the H$\alpha$ template to the individual radial velocity measurements and derive the systemic velocity for each star. We compute the uncertainty in the systemic velocity using Equation~9 of \citet{sesar12}, which assumes all distributions are Gaussian but accounts for the uncertainties in the individual velocity measurements and inherent uncertainties in the RR Lyrae magnitude--RV amplitude relation and template parameters. We take the systemic velocity to be the radial velocity of the fit template at phase 0.27, the canonical phase corresponding to zero pulsation velocity. The systemic velocity uncertainties are typically around 15--20 km~s$^{-1}$. Figure~\ref{fig:rvcurve} shows example template fits for two of the observed RR Lyrae stars (top panels) along with phase-folded, $R$-band light curves from PTF (bottom panels).


\subsection{Existing Catalina Sky Survey observations}

The TriAnd region defined in Section~2.1 is also covered by the Catalina Sky Survey \citep[CSS;][]{larson03}. One additional star in this area has a radial velocity measured by CSS follow-up efforts \citep{drake13} and we include this star in our sample of TriAnd RR Lyrae velocities. 90\% of the PTF RR Lyrae appear in the CSS sample, but there are many more CSS RR Lyrae in total. The CSS fields in this region have more observations and thus the completeness of RR Lyrae detection of the CSS is better than that of PTF. There are 142 RR Lyrae in CSS in the TriAnd volume, compared to 90 from PTF.


\section{Results I: the density and velocity distribution of TriAnd}\label{sec:resultsi}

We expect to find over-densities of RRLs in the TriAnd region consistent with the M giant population. In this section we compare the observed and the predicted number of RRLs and evaluate whether an over-density of RRLs is present. We then model the velocities of the observed RRLs and find that the RR Lyrae velocities appear to be consistent with a pure background (halo) velocity distribution.

\subsection{Density of RR Lyrae stars}

We estimate that the average CSS field in the TriAnd region has $\approx$300 observations. This is comparable to the average number of CSS observations along the SDSS Stripe 82 region, for which a previous study has identified nearly 100\% of the RR Lyrae stars in this region \citep{sesar10}. To estimate the completeness of RR Lyrae identification in CSS as a function of distance, we compare the number RR Lyrae in CSS relative to the number in the Stripe 82 sample in several distance bins ranging from 5 to 35 kpc. We find that between 15--21 kpc, CSS is 80\% complete for fields with several hundred observations; there are 77 RR Lyrae in this distance range, or, a completeness-corrected number $N_{\rm CSS} = 77/0.8 \approx 96$.

We compute the predicted number of RR Lyrae stars using the smooth halo density relation of Sesar et al. (2011),
\begin{align}
	\rho_{\rm RR}(r) &= \rho_0 \left(\frac{d_{\rm GC}}{r}\right)^\alpha\\
	\rho_0 &= 5.6~{\rm kpc}^{-3}\\
	\alpha &= 2.42
\end{align}
where $r$ is the Galactocentric radius \citep[with halo oblateness, $q=0.63$, taken into account][]{sesar11} and we fix the distance of the Sun to the Galactic center, $d_{\rm GC} = 8~{\rm kpc}$. We find the expected number of RR Lyrae stars to be  $N_{\rm exp} \approx 225$, over twice the completeness-corrected number of RRLs measured by the CSS. This is not surprising given the vast amount of substructure observed in halo RR Lyrae density \citep[e.g.,][]{sesar10, sesar13}, but this does indicate that the region around TriAnd1 is \emph{under-dense}, not over-dense, in RRLs. 

We now also compare the smooth model predictions for nearby regions of the sky with the number counts of RRLs from CSS. We find that in the adjacent region towards the Galactic anticenter, ($160^\circ < l < 220^\circ$, $-35^\circ < b < -15^\circ$), the single power-law model again vastly over-predicts the number of RRLs at distances between 15--21 kpc; we find a completeness corrected number of RR Lyrae in this volume $\approx$55 compared to the predicted number $\approx$190. In the opposite direction, the region adjacent and closer to the Galactic center (but slightly lower in $b$ due to survey footprint, $40^\circ < l < 160^\circ$, $-40^\circ < b < -20^\circ$) contains a completeness corrected estimate of $\approx$350 RRLs in the CSS and a predicted number $\approx$360. Again, in these high (absolute) latitude fields, many of the stars are halo stars where other authors have seen a large amount of substructure and variance to the RR Lyrae number counts \citep[e.g.,][]{sesar10, sesar13}.

\subsection{Analysis of the line-of-sight velocity distributions}\label{sec:velmodel}

To quantify the fraction of RR Lyrae in our sample that may belong to the TriAnd1 feature, we develop a statistical model and simultaneously fit the velocity profile and dispersion of the TriAnd M giants from \cite{sheffield14} with the RR Lyrae stars to assess how many RR Lyrae may be associated with the cold sequence. 

We model the line-of-sight velocities using a Gaussian mixture model: any individual velocity has some probability of being drawn from either an isotropic halo velocity distribution, or from a narrower TriAnd sequence:
\begin{align}
	p(v \given \boldsymbol{\theta}_{\rm TriAnd}, f) = 
	(1-f)\,\mathcal{N}(v \given v_{\rm h}, \sigma_{\rm h}^2) + f \, p(v \given \boldsymbol{\theta}_{\rm TriAnd})
\end{align}
where $v_{\rm h}$ and $\sigma_{\rm h}$ are the mean halo velocity and the halo velocity dispersion ---  fixed to $(v_{\rm h},\sigma_{\rm h})=(0, 106)~{\rm km~s}^{-1}$ \citep{brown10} --- $f$ is the fraction of stars of a given type that belong to TriAnd, $\boldsymbol{\theta}_{\rm TriAnd}$ are the parameters for describing the TriAnd velocity sequence, and $\mathcal{N}(x \given \mu, \sigma^2)$ is the normal distribution over $x$ with mean $\mu$ and variance $\sigma^2$. If we assume that the TriAnd stars are also Gaussian distributed around a linear velocity trend in Galactic longitude, $l$, we may write
\begin{align}
	p(v \given \boldsymbol{\theta}_{\rm TriAnd}) &= \mathcal{N}\left(v \given v_0 + (\dd v/\dd l)\,l, \, \sigma_v^2\right)\\
	\boldsymbol{\theta}_{\rm TriAnd} &= \left(v_0, \dd v/\dd l, \sigma_v^2\right)
\end{align}
where $v_0$ and $\dd v/\dd l$ are the zero-point and velocity gradient of the TriAnd feature, $\sigma_v$ is the intrinsic velocity dispersion of the TriAnd stars, and we assume the $l$ are measured perfectly.

Of course, the line-of-sight velocities are measured with finite uncertainties. If we assume the measured GSR velocity, $w$, is also Gaussian distributed with known uncertainty $\sigma_w$, we may write the joint likelihood over measured and true velocity as
\begin{align}
	p(w, v \given \boldsymbol{\theta}_{\rm TriAnd}, f) = \mathcal{N}(w \given v, \sigma_w^2) \, p(v \given \boldsymbol{\theta}_{\rm TriAnd}, f).
\end{align}
Marginalizing over the true velocity, $v$, we are left with the per-star likelihood of measured, GSR velocity, $w$, in terms of the mixture model parameters:
\begin{align}
	p(w \given \boldsymbol{\theta}_{\rm TriAnd}, f) &= \int p(w, v \given \boldsymbol{\theta}_{\rm TriAnd}, f) \, \dd v\\
	&= (1-f)\,\mathcal{N}(w \given v_{\rm h}, \sigma_w^2 + \sigma_{\rm h}^2) + f \, \mathcal{N}(w \given v_0 + (\dd v/\dd l)\,l, \sigma_w^2 + \sigma_v^2).
\end{align}
The M giant and RR Lyrae samples (with observed GSR velocities $w_{\rm MG}$ and $w_{\rm RRL}$, respectively) will have different TriAnd1 membership fractions, $f_{\rm MG}$ and $f_{\rm RRL}$. We may therefore write the full likelihood for both populations as
\begin{equation}
	p(v \given \boldsymbol{\theta}_{\rm TriAnd}, f_{\rm MG}, f_{\rm RRL}) = 
	p(v_{\rm MG} \given \boldsymbol{\theta}_{\rm TriAnd}, f_{\rm MG}) \, p(v_{\rm RRL} \given \boldsymbol{\theta}_{\rm TriAnd}, f_{\rm RRL})
\end{equation}
where the underlying model for the velocity sequence, parametrized by $\boldsymbol{\theta}_{\rm TriAnd}$, is the same. Assuming the velocity measurements are independent, the likelihood over all measured velocities, $\{w_i\}$, is the product of the individual (marginal) likelihoods:
\begin{equation}
	\mathcal{L} = p(\{w_i\} \given \boldsymbol{\theta}_{\rm TriAnd}, f_{\rm MG}, f_{\rm RRL}) = \prod_i p(w_i \given \boldsymbol{\theta}_{\rm TriAnd}, f_{\rm MG}, f_{\rm RRL}).
\end{equation}
The posterior probability is then 
\begin{equation}
	p(\boldsymbol{\theta}_{\rm TriAnd}, f_{\rm MG}, f_{\rm RRL} \given \{w_i\}) = \frac{1}{\mathcal{Z}} \, p(\{w_i\} \given \boldsymbol{\theta}_{\rm TriAnd}, f_{\rm MG}, f_{\rm RRL}) \, p(\boldsymbol{\theta}_{\rm TriAnd}) \, p(f_{\rm MG})\, p(f_{\rm RRL}) \label{eq:post}
\end{equation}
where the evidence integral, $\mathcal{Z}$, is a constant that depends only on the data, $p(\boldsymbol{\theta}_{\rm TriAnd})$ is the prior probability over the TriAnd velocity model parameters, and $p(f_{\rm MG})$ and $p(f_{\rm RRL})$ are the prior probabilities over the fraction of M giants and RRLs, respectively, that are halo (field) stars. We assume a uniform prior for the zero-point of the TriAnd velocity sequence, $v_0$, over the range $(-200,200)$~km~s$^{-1}$, a scale-invariant (logarithmic) prior over the velocity dispersion of TriAnd, $\sigma_v^2$, and assume uniform priors over each membership fraction over the range (0,1).

\begin{figure}[p]
\begin{center}
\includegraphics[width=\columnwidth]{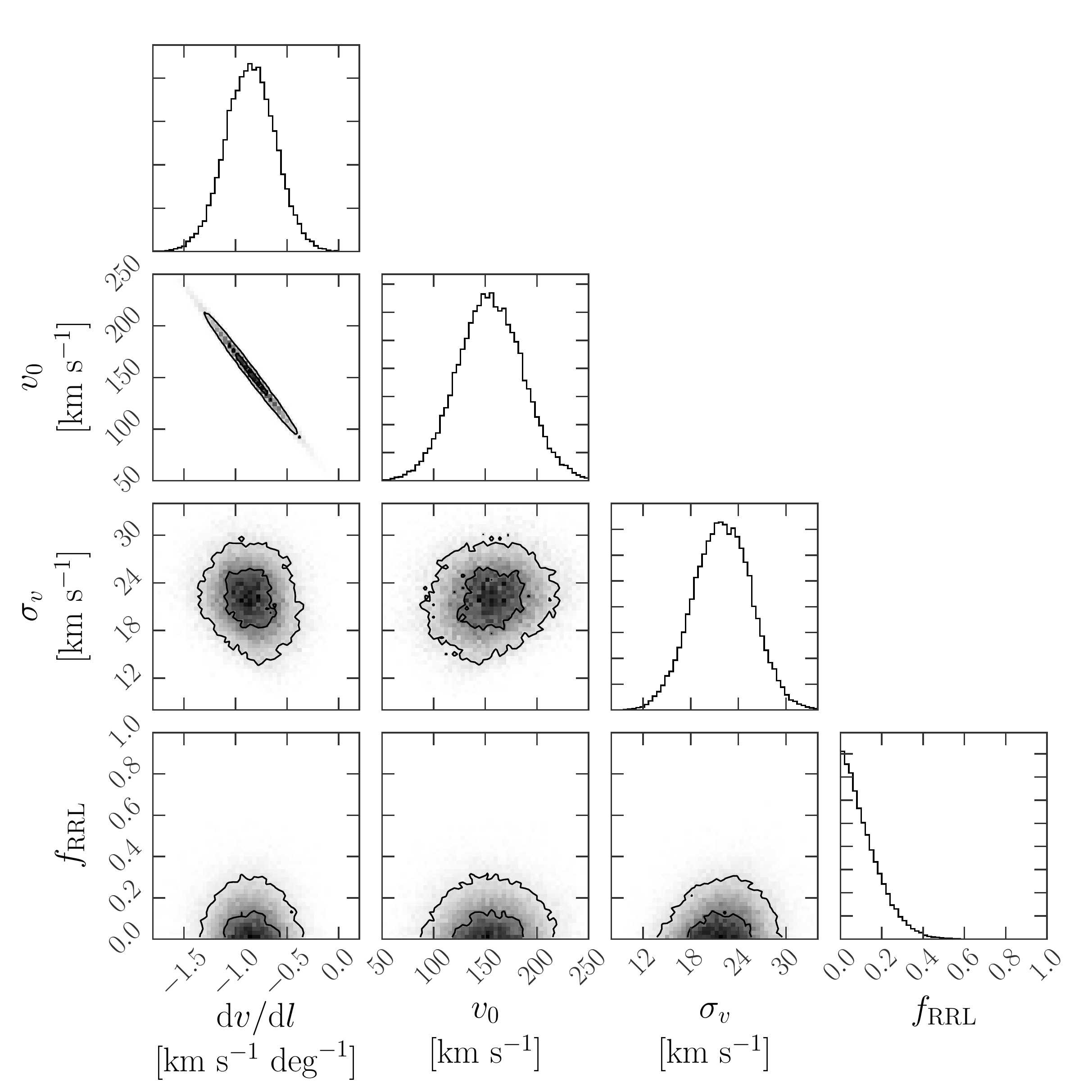}
\caption{Projections of the TriAnd velocity model posterior probability distribution (Equation~\ref{eq:post}). Contours show (approximate) 68\% and 95\% confidence regions in each projection.  \label{fig:posterior}}
\end{center}
\end{figure}

We use an ensemble Markov Chain Monte Carlo (MCMC) algorithm \citep{goodman10} implemented in \texttt{Python} \citep{foreman-mackey13} to draw samples from the posterior probability distribution defined above (Equation~\ref{eq:post}). The algorithm uses an ensemble of ``walkers'' that naturally adapt to the geometry of the parameter-space; we use 32 walkers in this analysis. We burn-in the walkers for 256 steps starting from randomly drawn initial conditions (sampled from the priors) and then re-initialize the walkers from their final burn-in positions and run for another 4096 steps. Figure~\ref{fig:posterior} shows contour-plot projections of the samples from the posterior. We find that the TriAnd velocity sequence has a velocity dispersion, gradient, and zero-point of 
\begin{align}
	\sigma_v &= 22.0 \pm 3.6~{\rm km}~{\rm s}^{-1}\\
	\dd v / \dd l &= -0.86 \pm 0.24~{\rm km}~{\rm s}^{-1}~{\rm deg}^{-1} \\
	v_0 &= 155 \pm 32~{\rm km}~{\rm s}^{-1}
\end{align}
but with significant correlation between $v_0$ and $\dd v / \dd l$. 

For each star (M giant and RR Lyrae), we compute the the posterior probability that the star is a member of the TriAnd velocity sequence \citep{dfmMixtureModel14}. To do this, we make use of an additional binary, nuisance parameter, $q$ --- when $q=0$, the star is part of the halo population, and when $q=1$, the star is a member of TriAnd. The conditional probability that a given star is a member of TriAnd (e.g., $q=1$) is
\begin{align}
	p(q=1 \given \boldsymbol{\theta}_{\rm TriAnd}, f, w) = \frac{f \, \mathcal{N}(w \given v_0 + (\dd v/\dd l)\,l, \sigma_w^2 + \sigma_v^2)}{p(w \given \boldsymbol{\theta}_{\rm TriAnd}, f)}
\end{align}
and the posterior probability is just the integral of this expression over all parameters. This integral can be approximated with a Monte Carlo integral using the posterior samples generated with MCMC,
\begin{align}
	p(q=1 \given w) &= \int p(q=1 \given \boldsymbol{\theta}_{\rm TriAnd}, f, w) \, p(\boldsymbol{\theta}_{\rm TriAnd} , f\given w) \, \dd f \, \dd \boldsymbol{\theta}_{\rm TriAnd}\\
	P_{\rm TriAnd} &\approx \frac{1}{N} \sum_n^N \, p(q=1 \given \boldsymbol{\theta}_{\rm TriAnd}^{(n)}, f^{(n)}, w).
\end{align}
Figure~\ref{fig:data} (top panel) shows the M giant stars with $P_{\rm TriAnd} > 0.8$ in black, and M giants identified as being background halo stars in grey. 


\subsection{Estimated Ratio of RR Lyraes to M giants in TriAnd: $f_{\rm RR:MG}$}

\begin{figure}[p]
\begin{center}
\includegraphics[width=\columnwidth]{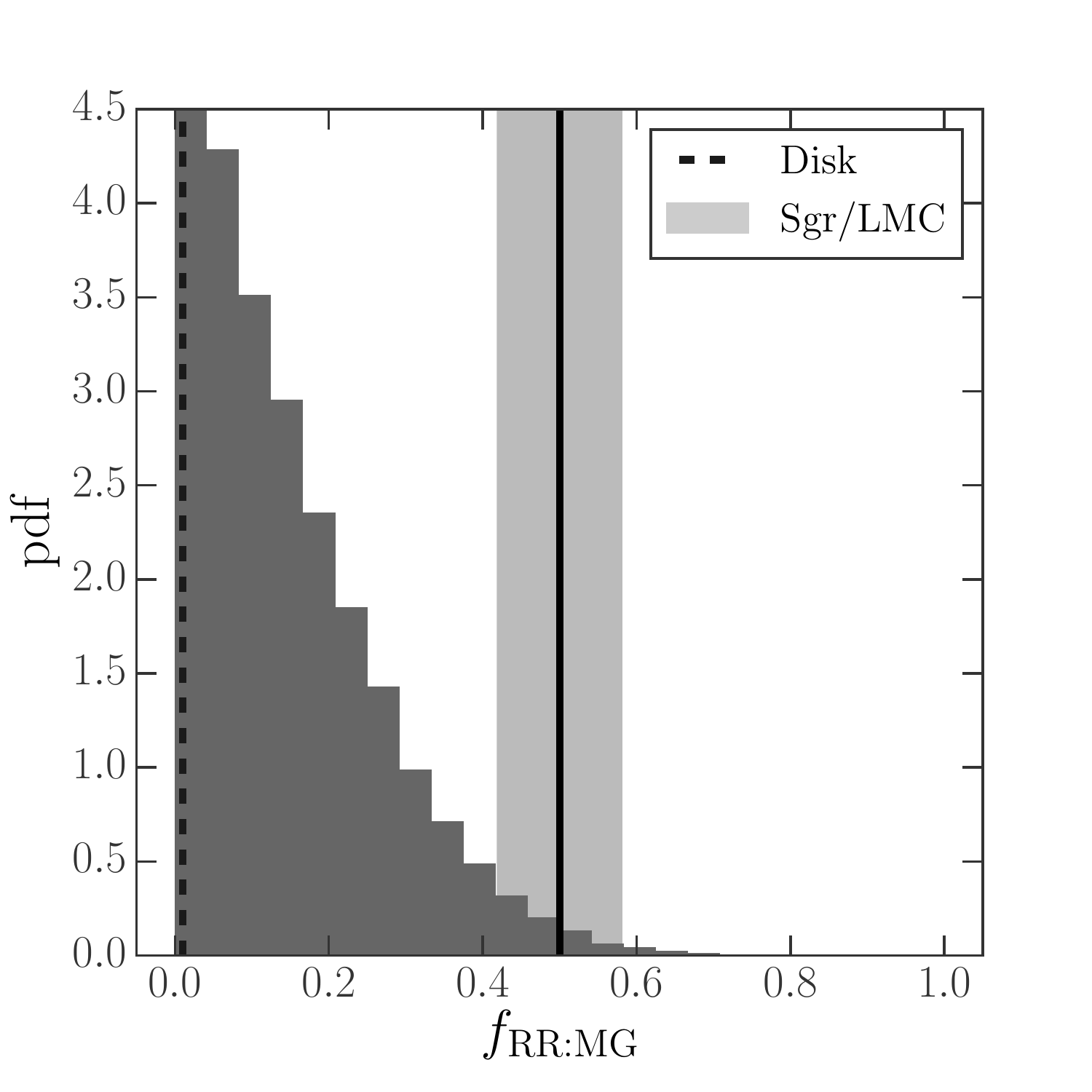}
\caption{Binned samples from the marginal posterior probability over the number ratio of RR Lyrae to M giants in TriAnd (histogram). Dashed, vertical line shows the value of the ratio typical to the Milky Way disk ($<$1\%). Solid, vertical line and shaded rectangle indicate the approximate mean and dispersion of $f_{\rm RR:MG}$ for counts around both Sgr and the LMC (see text for exact numbers). \label{fig:frrmg}}
\end{center}
\end{figure}

We simultaneously constrain the fraction of RR Lyrae that may belong to the cold TriAnd velocity sequence with our velocity model. Figure~\ref{fig:posterior} (bottom right) shows the marginal posterior distribution over $f_{\rm RRL}$; the MAP value of $f_{\rm RRL} = 0.004$, and the values $f_{\rm RRL} = (0.183, 0.38)$ contain $68\%$ and $95\%$ of the probability mass, respectively. We use the marginal posterior over $f_{\rm RRL}$, the total (completeness-corrected) number of RR Lyrae in this region, $N_{\rm CSS}$, and the number of M giants identified by \cite{sheffield14} to be in TriAnd1, $N_{\rm MG}=74$, to compute the distribution over the ratio of the number of RR Lyrae to the number of M giants in TriAnd, $f_{\rm RR:MG} = f_{\rm RRL} \, N_{\rm CSS} / N_{\rm MG}$. Figure~\ref{fig:frrmg} shows our posterior constraints on this ratio.

\begin{figure}[p]
\begin{center}
\includegraphics[width=0.98\columnwidth]{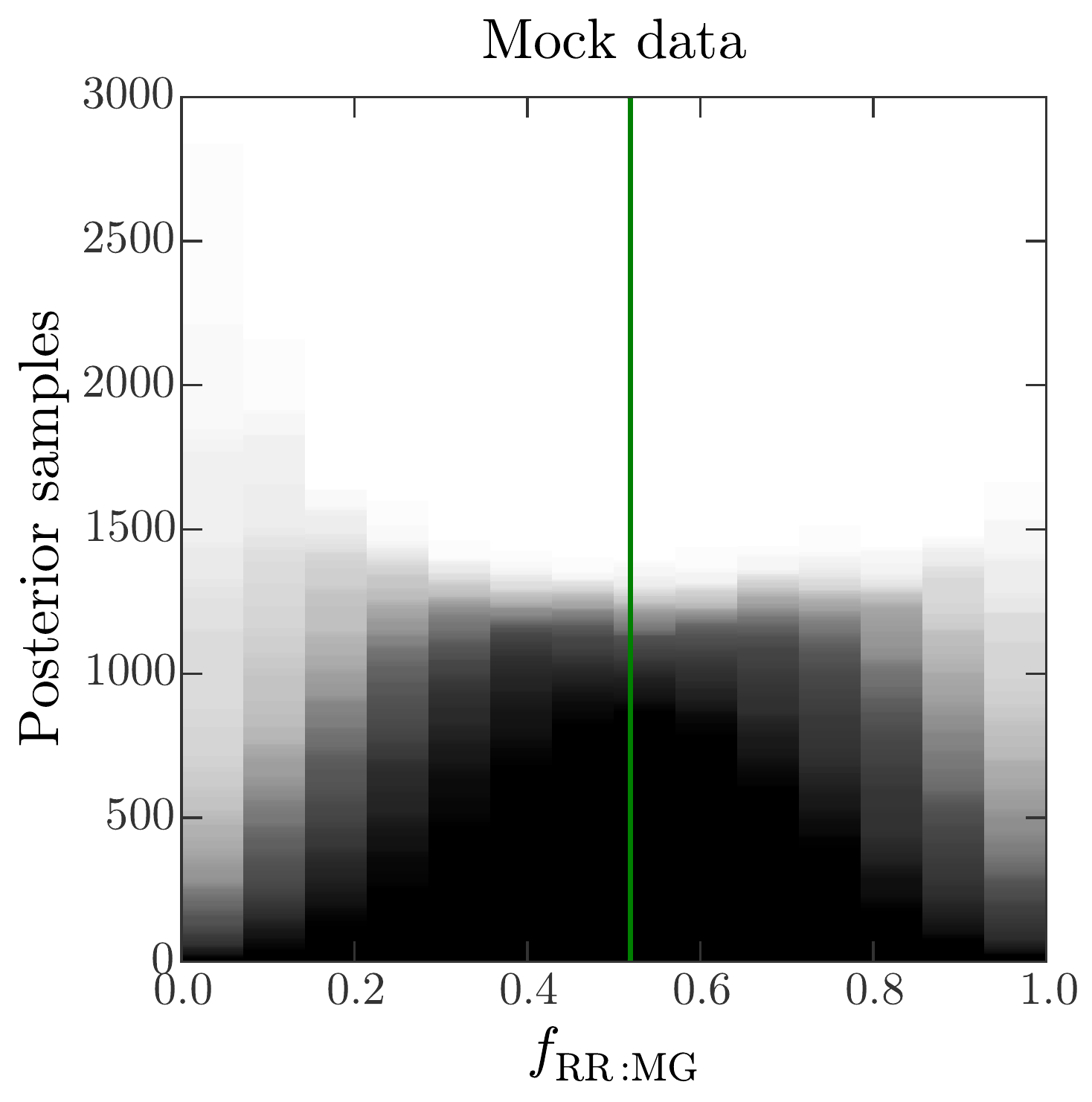}
\caption{Histograms of samples from 1024 individual posterior probability distributions (plotted with transparency) over the number ratio of RR Lyrae to M giants in mock data sets of RR Lyrae velocities. The RR Lyrae velocities are drawn from either an isotropic halo distribution or a thin, linear velocity sequence with known properties and intrinsic velocity dispersion set to 20 km~s$^{-1}$. Each velocity is observed with an uncertainty drawn from a uniform distribution over the range 15--20 km~s$^{-1}$. Vertical (green) line shows the true, input value of the ratio. \label{fig:mockdata}}
\end{center}
\end{figure}

To validate our model and test whether we would expect to detect a nonzero fraction of RR Lyrae stars associated TriAnd with our data, we generate mock RR Lyrae velocity data and use the model to infer $f_{\rm RR:MG}$ for many trials. We assume that there are 100 RR Lyrae in total and 75 M giants in our mock TriAnd region, but only ``observe'' 20 RR Lyrae stars assuming 40\% of the stars are sampled from a low-dispersion, linear velocity sequence with an intrinsic scatter set to 20 km~s$^{-1}$. These numbers produce a true RR Lyrae to M giant fraction, $f_{\rm RR:MG}$, equal to $\approx$0.5, which represents a \emph{lower bound} for what we would expect for accreted stellar populations (see Section~\ref{sec:resultsii}). The rest of the stars are drawn from the same isotropic halo distribution used above. For each mock RR Lyrae star, we observe the velocity with an uncertainty drawn from a uniform distribution over the range 15--20 km~s$^{-1}$. For this test, we assume we know the true TriAnd velocity trend and only consider a single free parameter: the fraction of RR Lyrae that belong to the cold sequence, which we convert to $f_{\rm RR:MG}$ using the numbers above. We perform 1024 trials in this test. Figure~\ref{fig:mockdata} shows histograms of posterior samples over $f_{\rm RR:MG}$ for each of the 1024 trials, plotted with transparency. The green line indicates the true fraction in the mock dataset. The posterior distribution over $f_{\rm RR:MG}$ for any individual trial with a true ratio of $f_{\rm RR:MG} \approx 0.5$ generally recovers a fraction between 0.3--0.7 and is almost never peaked towards zero.

\section{Results II: Implications for the origin of the TriAnd1 and TriAnd2 substructures}\label{sec:resultsii}

\subsection{Accreted satellite?}

The implications of {\it not} being able to easily find RR Lyrae associated with TriAnd could be very interesting. 
{\it All} dwarf galaxies that are close enough to have been surveyed for RR Lyrae are found to contain at least a few \citep{clementini10}. However, only the largest satellites (e.g. the Small and Large Magellanic Clouds -- SMC, LMC --  and the Sagittarius dwarf spheroidal galaxy -- Sgr) are known to contain M giants.
This systematic difference between low- and high-luminosity dwarf galaxies is easily explained as a consequence of the age and metallity distributions of their stellar populations.
Single age populations old enough to containing stable, helium core-burning stars have to be less metal rich than ${\rm [Fe/H]}\approx -1$ for their horizontal branches to extend beyond the instability strip that RR Lyrae stars inhabit.
Conversely, populations have to be more metal rich than ${\rm [Fe/H]}\approx -1$ for red-giant-stars to evolve through spectral types as late as M.  
Only the LMC, SMC, and Sgr have metallicity distributions that are both metal-rich enough to contain M giants  yet with a sufficiently populated tail to lower metallicities that they also contain significant numbers of RR Lyrae \citep[e.g.][]{chou07,carrera08,kirby11}. 
\citep[Note, while the ages and metallicity estimates from isochrone fitting in][are inconsistent with RR-Lyrae populations, these estimates were made under the {\it assumption} that the main-sequence detections corresponded to the overdensities seen in the M giant populations and hence do not themselves rule out the presence of RR Lyrae in these structures.]{sheffield14}

If the TriAnd structures did form from the disruption of a satellite, then the presence of significant M giants in them suggests that the progenitor was likely to have stellar populations like the LMC, SMC, and Sgr.
Hence, we can use the ratio of the numbers of RR Lyrae and M giants, $f_{\rm RR:MG}$, in these satellites as a guide for how many RR Lyrae we might find also associated with these structures if they were likely accreted from a large satellite.
 
\begin{figure}[t]
\begin{center}
\includegraphics[width=1.0\columnwidth]{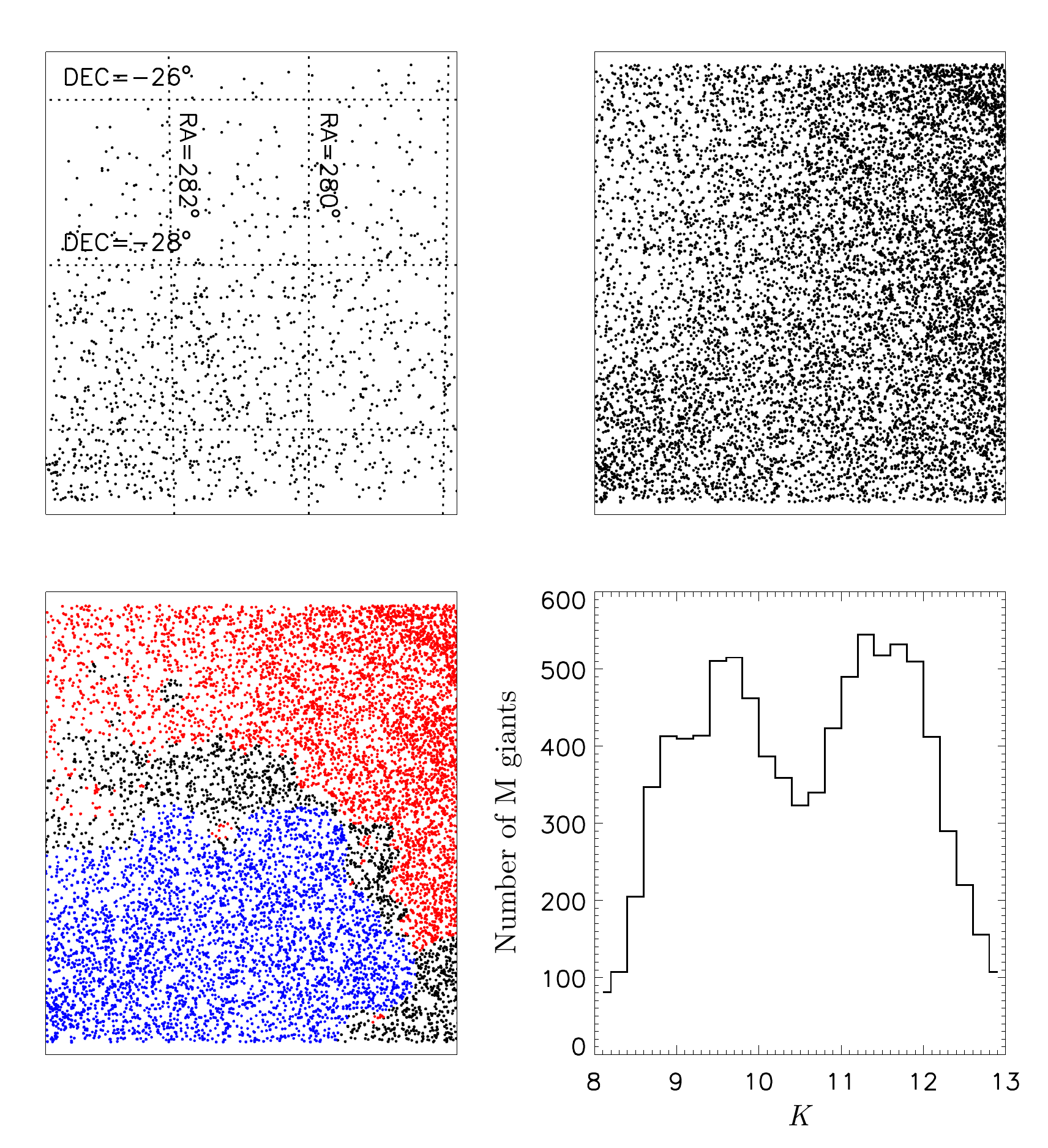}
\caption{\label{fig:sgr} 
Top left panel shows the position of RR Lyrae in the SAG plate of the \citet{cseresnjes00} catalogue.
Top right panel shows M giants selected from 2MASS in the same region (see text for details).
Bottom left panel repeats the top right panel with stars color coded in red that have ${\rm E}(B-V)>0.2$ and blue to show fields where stars were counted to calculate $f_{\rm RR:MG}$.
The bottom right panel is a histogram of the apparent magnitude distribution of the M giants in all the blue fields combined.
}
\end{center}
\end{figure}

Figure \ref{fig:sgr} summarizes how we estimated $f_{\rm RR:MG}$ for Sgr.
The top left-hand panel shows the positions of RR Lyrae in the SAG plate of the catalogue of \citet{cseresnjes00}, which maps the outer structure of Sgr and its debris at Galactic latitudes above the main progenitor, $-14^\circ < l < -5^\circ$.
Only those with $B$-magnitudes greater than 18.5 are shown, which selects Sgr stars with only mild contamination from the bulge \citep[see][Figure 3]{cseresnjes00}.
The top right-hand panel shows the  M giants in same region selected from the 2MASS catalogue \citep[following][]{majewski03} in the same areas.
The M giants were chosen within appropriate color limits \citep[after reddening correction and to match the selection criteria, $0.9<(J-K_s)<1.3$, employed for the TriAnd sample in][]{sheffield14} and chosing (extinction-correction) magnitudes ($8 < K_s <13$)  that would encompass the full range possible given the distance to Sgr and the known spread in M-giant luminosities.
The bottom left-hand panel repeats the middle panel, but with M giants with reddening-estimates ${\rm E}(B-V)>0.2$ highlighted in red.
Clearly, the red points in this panel cluster around the apparently ``empty'' regions of the plot, and the clumpy nature of the M giant distribution compared to the RR Lyrae can be attributed to reddening effects.
To address this incompleteness we placed a grid of overlapping fields of radius 0.5$^\circ$ over the area and made separate estimates of the number of M giants and RR-Lyrae for only those that contained no M giant with ${\rm E}(B-V)>0.2$ (highlighted by blue points in the same panel).

The bottom right-panel of Figure \ref{fig:sgr}, which shows the magnitude-distributions for M giants for the fields highlighted in blue in the bottom left-hand panel,  illustrates how the final estimates for $f_{\rm RR:MG}$ were made. 
For the M giants, the broad spread in absolute magnitudes causes significant overlap in the apparent magnitude distributions of the (brigter, $K_s \sim 9-10$) bulge stars and (fainter, $K_s \sim 11-12$ ) Sgr stars, so in this case the contribution of Sgr stars in a given field was estimated by fitting two Gaussians (with means, dispersions, and amplitudes as free parameters) to the full distribution and estimating the number of Sgr stars in the combined fields from the ratio of the the amplitudes.
Overall, we found a total of 2319 RR Lyrae and  4770 M giants across all fields associated with Sgr, suggesting $f_{\rm RR:MG}\approx 0.5$.
The ratio in the 41 (overlapping) individual fields had a mean and dispersion in $f_{\rm RR:MG}$ of 0.48 and 0.088 (and varied across the full range 0.31-0.69).
The individual fields had $\approx$ 50--100 and $\approx$ 100--200 in them respectively, suggesting that fluctuations in counts should lead to a dispersion in $f_{\rm RR:MG}$ of order $0.02$. 
We interpret the larger measured dispersion as due to a combination of possible factors including intrinsic variations in stellar populations, uncertainties in the counts themselves in these crowded fields and Gaussians not being an appropriate form for the fitting function.

\begin{figure}[t]
\begin{center}
\includegraphics[width=1.0\columnwidth]{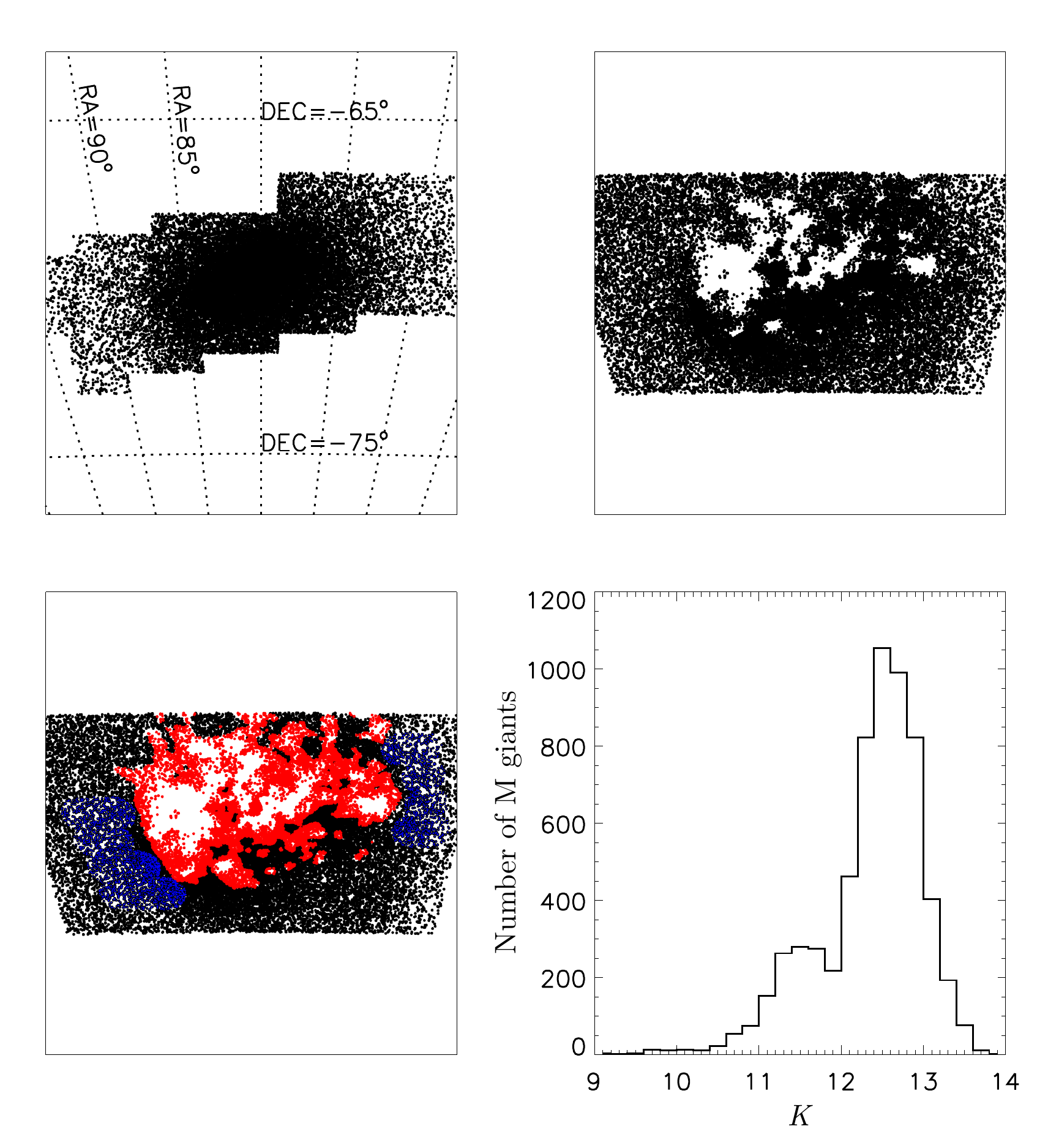}
\caption{\label{fig:lmc} 
Top left panel shows the position of RR Lyrae in LMC region in the \citet{soszynski09} catalogue.
Top right panel shows M giants selected from 2MASS in the same region (see text for details).
Bottom left panel repeats the top right panel with stars color coded in red that have ${\rm E}(B-V)>0.2$ and blue to show fields where stars were counted to calculate $f_{\rm RR:MG}$.
The bottom right panel is a histogram of the apparent magnitude distribution of the M giants in all the blue fields combined.
}
\end{center}
\end{figure}

Figure \ref{fig:lmc} repeats the same plots as Figure \ref{fig:sgr} for the LMC, where the RR Lyrae counts come from the catalogue of the OGLE team \citep{soszynski09}. The same approach as for Sgr was used to find an overall number of 2686 RR Lyrae and 5210 M giants in the fields we examined, suggesting $f_{\rm RR:MG}\approx 0.5$.
The ratio in the 18 (overlapping) individual fields had a mean and dispersion in $f_{\rm RR:MG}$ of 0.52 and 0.086 
(and varied across the full range 0.39-0.69).
The individual fields had $\approx$ 100--200 RR Lyrae and $\approx$ 200--400 M giants in them. 
We again interpret the large dispersion as due to a combination of the same possible factors.

The estimates and ranges for $f_{\rm RR:MG}$ from our analyses of Sgr and the LMC are plotted as lines and shaded regions on top of the PDF for $f_{\rm RR:MG}$ generated from our data in the TriAnd region in Figure \ref{fig:frrmg}.
While an Sgr- or LMC-like population cannot yet be absolutely ruled out, the data suggests that the TriAnd structures ($f_{\rm RR:MG} < 0.38$ at 95\% confidence) have a much lower ratio of RR Lyrae to M giants than these objects ($f_{\rm RR:MG} \approx 0.5 \pm 0.1$), and very unlike any of the other known satellites which are not known to contain M giants at all.

Of course, the disrupted object might not have populations like the Sgr or LMC.
However a lower luminosity or more ancient satellite would only have older and/or lower metallicity populations, suggesting a larger fraction of RR Lyrae.
A more massive satellite might be higher metallicity overall and contain a smaller fraction of RR Lyrae, but then would be likely to give TriAnd a larger dispersion than observed and leave many other signatures of its recent disruption in the disk and the halo.
Overall, we conclude that the lack of RR Lyrae stars contributing the cold velocity sequence outlined by M giants in the region   calls into question the interpretation of the origin of TriAnd structures as satellite debris.

\subsection{Kicked-out disk?}

In comparison to the Galactic satellites, the Galactic disk is more metal rich and its metal-poor tail is much more poorly populated \citep{schlesinger12} --- hence it contains a much higher fraction of M giants. 
The observed number density for RR Lyrae at the disk midplane around the Sun is $\approx$ 5--10 per kpc$^3$ \citep[predominantly from the thick disk, see][]{layden95,amrose01} while  the number density of M giants predicted by the {\it Galaxia} stellar populations code is $\approx$ 4000 per kpc$^3$ \citep{sharma11}, suggesting a number ratio of less than 1\%.
Hence, the apparent lack of RR Lyrae associated with the TriAnd velocity sequences could be indicative of these populations originating in our Galactic disk.
The ages and metallicities of the M giants in these sequences indicated by isochrone fitting  in \citet{sheffield14} (ages of 6-8Gyrs {\it vs} 10-12 Gyrs, and [Fe/H] $\sim$ -0.75 {\it vs} - 0.85 for TriAnd1 and TriAnd2 respectively)  are loosely consistent with an extrapolation of the scales and gradients of metallicities seen at smaller radii (3-15kpc) in the disk, along with the models of age distributions that these observations suggest \citep{hayden15}.
However, there is not yet enough known about stellar populations at these extreme radii for this extrapolation to be conclusive in itself.

We conclude that the stellar populations of TriAnd1 and TriAnd2  suggest these substructures are likely due to a perturbation kicking stars out of the disk rather than a satellite accretion event.
This formation scenario has not been considered until very recently \citep[][]{xu15} due to the large Galactocentric radius (30--40 kpc) and position below the disk plane (5--10 kpc) of these features. 
In the next section we examine whether coherent stellar rings can be kicked to such extreme locations through encounters with known satellites.

\section{Results III: the creation of extreme disk populations}\label{sec:resultsiii}

In order to investigate the plausibility of creating TriAnd1 and TriAnd2 with a dynamical disturbance to the Galactic disk, we performed N-body simulations of three-component (disk, bulge and dark-matter halo) Milky-Way-like galaxies being perturbed by a satellite with the mass and orbital characteristics of the Sagittarius dwarf galaxy. The initial conditions were generated using the rejection method \citep{kuijken94} to sample the distribution functions of each components as in \citet{widrow08}. This method is reliable and shown to produce more stable galaxies \citep{kazantzidis04} than methods relying on moments of the collisionless Boltzmann equation which have typically been used in the literature \citep{hernquist93}. The simulations were evolved with the Gadget-3 treecode \citep{springel01,springel05}.

There exists prior work in the literature on the impact of MW satellites (LMC or Sgr) on our Galactic disk\citep[e.g.][]{weinberg95,bailin03}. In particular, \citep{purcell11} qualitatively show that the impact of a dark matter halo of the size that could have hosted the progenitor of Sgr could plausibly produce the Monoceros Ring or the Giant Anticenter Stellar Structure \citep{yanny03,crane03} as well as the north/south velocity asymmetries in the Galactic disk that have been seen in the RAVE and SEGUE surveys \citep{widrow12,gomez13,williams13}. As a starting point to our investigation, we have re-run a similar N-body simulation to \cite{purcell11}. The dark matter halo of the MW galaxy is given by an NFW halo with a virial mass of $M_{vir}=10^{12}~\msun$ and a scale radius of $r_{s}=14 ~ {\rm kpc}$. The disk has a mass of $3.7 \, \times 10^{10}~\msun$, an exponential scale length of $2.8 ~ {\rm kpc}$ and a vertical scale height of $0.4 ~ {\rm kpc}$ and a cut-off radius at $\approx 50$ kpc. The bulge has a mass of $9.4 \times 10^{9}~\msun$ and a S\'ersic profile with index $n=1.28$ and effective radius of $0.56~{\rm kpc}$. The Sagittarius analogue is modeled with a virial mass of $M_{vir}=10^{11}~\msun$ and scale length $r_{s}=6.5 ~ {\rm kpc}$. The satellite is launched at 80 kpc from the Galactic centre from the plane of the MW on a vertical trajectory with a velocity vector $\mathbf{v}=(v_x,v_y,v_z)=(0,0,-80) ~ {\rm km~ s}^{-1}$ towards the Galactic North. At those distances, the halo of the satellite galaxy will have experienced significant mass-loss due to stripping so we artificially take into account this effect by truncating the density profile at the tidal radius $r_{t}=30~{\rm kpc}$. This makes the satellite mass a factor of $\approx$3 smaller than the initial virial mass originally set. The simulation is run for 2.4 Gyr and the Sgr dwarf reaches present-day configuration at $\approx$ 2.3 Gyr. Our simulation followed the evolution of $\approx 7 \times 10^{7}$ particles ($N_{\rm disk}=6\times10^{6}$, $N_{\rm bulge}=2 \times 10^6$, $N_{\rm halo}=4\times10^{7}$, $N_{\rm Sgr}=4.2\times 10^{6}$). The dark matter particle mass was $\approx$ 4 times larger than disk particles. The initial disk model is tested for stability by running it for a period of 2 Gyr. Apart from the appearances of transient flocculent spirals the disk showed no variations in its density profile and showed no appreciable disk thickening due to interactions with the more massive MW surrounding dark matter particles.

\begin{figure}[p]
\begin{center}
\includegraphics[width=1.\columnwidth]{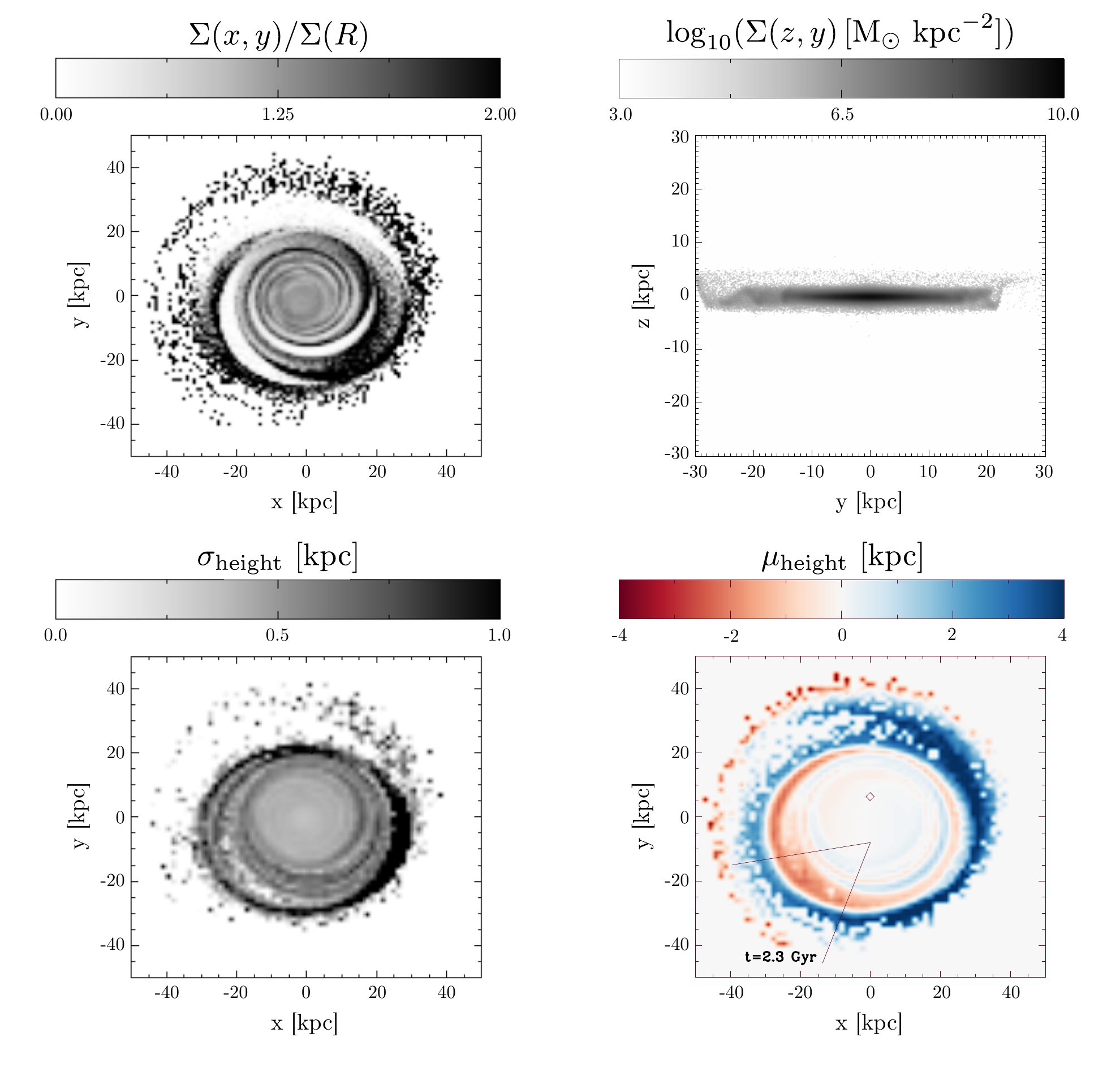}
\caption{
\label{fig:disk}
Top left-hand panel shows the density contrast map of our simulated galactic disk at 2.3 Gyr  ($\Sigma(x,y)/\Sigma(R)$, where $\Sigma(R)$ is the azimuthally averaged density and $\Sigma(x,y)$ is the local surface density evaluated at position $(x,y)$). Top right-hand panel shows an edge-on view of the disk --- vertical oscillations are most apparent towards negative $y$.
The bottom panels show the mean position (right) and dispersion (left) of particles perpendicular to the galactic disk.
The diamond in the bottom right panel indicates the instantaneous position of the perturbing satellite (projected) and the lines indicate position of the Sun and the range of Galactic longitude occupied by the TriAnd Clouds.}
\end{center}
\end{figure}

Figure \ref{fig:disk} illustrates the results at the end point of 2.3 Gyr of evolution in our simulation.
The top-left, bottom-left, and bottom-right panels show two-dimensional maps of the over-density, average $z$-position, and  $z$-dispersion for particles in the disk component. Top-right shows an edge-on view of the simulated disk.
Outward-propagating ring-like features are a natural product of the interaction,
with several rings produced during the course of the simulation, over-dense by a factor of $\approx 2$ relative to the rest of the disk and spaced by several kpc.

The bottom-right panel demonstrates that these over-densities oscillate in $z$ above and below the plane, with the amplitude of the oscillation increasing with $R$ to as much $\approx 4$ kpc at $R=30~{\rm kpc}$. The thickness of the disk --- as indicated by the dispersion $\approx$ 0.5-1.0 kpc in the top-right and bottom-left panel --- also varies. An observer at 8 kpc along the $y$-axis in our simulation would clearly see two of these oscillations as rings several kpc {\it below} the galactic disk at roughly 18 and 40 kpc beyond their radius. (A third, thinner ring appears at $\approx$ 10 kpc distance, but its significance is less apparent.) The properties of these simulated structures are reminiscent of the GASS and TriAnd clouds (albeit with smaller offsets from the disk plane) and support the hypothesis that these structures could plausibly be caused by the same interaction. 

There are many things that could explain why the observed oscillations are larger than those seen in the simulations.
The initial conditions in our simulation --- with the satellite artificially truncated at its pericentric tidal radius and the disk in equilibrium --- neglects the first tidal interactions between the two.
Moreover, Sgr is not the only object in orbit around the Milky Way capable of significantly perturbing the disk: the effect of the LMC should also be taken into account.
Further, the disk itself is already known to be warped; whether this warp is a signature of past or present interactions, its presence  will affect any further reactions to ongoing perturbations.
Finally, the properties of neither the stellar disk nor the Galactic potential are well-known in this region and could significantly affect the response to any interaction.

The above list of possible explanations is challenging to investigate. At the same time, the presence of this disk population at such extreme Galactic radius and distance from the Galactic plane are challenging to reproduce, which suggests that their existence could provide strong tests of the various hypotheses.

\section{Conclusions and Future Work}\label{sec:conclusions}


By looking at $f_{\rm RR:MG}$ --- the ratio of the number of RR Lyrae to M giant stars --- in the TriAnd1 structure, we have shown that the stellar populations in this M giant over-density in the halo appear more like those in the disk of our Milky Way than 
in any of the Galactic satellites.
We have also confirmed, using N-body simulations, that the impact of a large satellite can plausibly generate concentric rings of stars in the outskirts of the Galactic disk propagating outwards from the Galactic center (albeit with lower amplitude). 
We conclude that TriAnd1 and TriAnd2 could plausibly be two such propagating stellar rings. 
As such, they would represent the first disk populations seen at such large $R$ and $Z$ around the Milky Way.
Their existence provides evidence for stars currently being {\it kicked-out} of the disk, and strong support for this formation mechanism contributing a significant fraction of our stellar halo.

This work bolsters the tentative connections proposed by \citet{xu15} between various over-densities seen in the star counts in the outer disk and the local velocity asymmetries in the disk, which they propose to represent the collective response to an external perturbation.
 Further work on abundance pattern \citep[as started in][]{chou11} and stellar populations could help confirm or rule out this scenario. 
A similar test of number ratios could be used to further examine whether other over-densities at smaller $R$ and $Z$ \citep[such as GASS -- the Giant Anticenter Stellar Stream --  also known as the Monoceros Ring;][]{rochapinto03,ibata03} could be the inner rings excited  by the same disturbance to the Galactic disk.

Our results also motivate more careful and complete numerical explorations of dynamical interactions in order to understand under what circumstances disk material can be kicked coherently to such extreme locations. 
Analogous work on the signatures in outskirts of the gaseous Galactic disk \citep{chakrabarti11} has recently led to the identification of a new candidate dwarf satellite at 90 kpc from the Milky Way \citep{chakrabarti15}.
Hence we anticipate that the mere existence of the TriAnd structures could place strong constraints on such interactions and the potential of the Milky Way.

\acknowledgements
This material is based upon work supported by the National Science Foundation under grant number NSF-AST 1312196.
APW is supported by a National Science Foundation Graduate Research Fellowship under Grant No.\ 11-44155 and acknowledges support from the 2014 Sigma Xi Grants-in-Aid of Research (GIAR) program. This research made use of Astropy, a community-developed core \texttt{Python} package for Astronomy \citep{astropy13}. This work relied on Columbia University's \emph{Hotfoot} and \emph{Yeti} compute clusters, and we acknowledge the Columbia HPC support staff for assistance. The authors acknowledge the Texas Advanced Computing Center (TACC) at The University of Texas at Austin for providing HPC resources that have contributed to the research results reported within this paper. 
This work is based on observations obtained at the MDM Observatory, operated by Dartmouth College, Columbia University, Ohio State University, Ohio University, and the University of Michigan. We thank Jules Halpern for advice and discussion about spectroscopic data reduction, Gisella Clementini for consultation on the RR Lyrae population in dwarf satellites, and Greg Bryan for access to an allocation on \emph{Stampede}. CPFL thanks Facundo Gomez for useful discussions.
Many thanks to the ``Clouds'' team (Steve Majewski, Ting Li, Jennifer and  Jeff Carlin) for the ongoing collaboration from which this work developed.
This article is based on observations obtained with the Samuel Oschin
Telescope as part of the Palomar Transient Factory project, a scientific
collaboration between the California Institute of Technology, Columbia
University, Las Cumbres Observatory, the Lawrence Berkeley National
Laboratory, the National Energy Research Scientific Computing Center,
the University of Oxford, and the Weizmann Institute of Science. We wish
to acknowledge builders of the PTF photometric pipeline, Judith G. Cohen
for leading the RR Lyrae effort within PTF, and Shrinivas R. Kulkarni
for leading the PTF project.

\bibliographystyle{apj}
\bibliography{refs_adrian,refs_lsb_disk,refs_chervin,refs_extra,refs_revisions}
\clearpage

\end{document}